\documentclass[12pt]{article}

\def\aj{AJ}%
\def\apj{ApJ}%
\def\apjl{ApJ}%
\def\prd{Phys.~Rev.~D}%
\def\pasp{PASP}%
\def\nat{Nature}%
\usepackage{scicite}

\usepackage{epsfig}
\usepackage{times}
\usepackage{graphicx}
\usepackage{url}

\newenvironment{sciabstract}{%
\begin{quote} \bf}
{\end{quote}}

\newcounter{lastnote}

\title{Relativistic Spin Precession in the Double Pulsar}

\author
{Rene P. Breton,$^{1\ast}$ Victoria M. Kaspi,$^{1}$ Michael Kramer,$^{2}$\\
Maura A. McLaughlin,$^{3,4}$ Maxim Lyutikov,$^{5}$ Scott M. Ransom,$^{6}$\\
Ingrid H. Stairs,$^{7}$ Robert D. Ferdman,$^{7,8}$\\
Fernando Camilo$^{9}$, Andrea Possenti$^{10}$\\
\\
\normalsize{$^{1}$Department of Physics, McGill University, Montreal, QC H3A 2T8, Canada}\\
\normalsize{$^{2}$Jodrell Bank Observatory, University of Manchester, Manchester, M13 9PL, UK}\\
\normalsize{$^{3}$Department of Physics, West Virginia University, Morgantown, WV 26506, USA}\\
\normalsize{$^{4}$National Radio Astronomy Observatory, Green Bank, WV 24944, USA}\\
\normalsize{$^{5}$Department of Physics, Purdue University, West Lafayette, IN 47907, USA}\\
\normalsize{$^{6}$National Radio Astronomy Observatory, Charlottesville, VA 22903, USA}\\
\normalsize{$^{7}$Department of Physics and Astronomy, University of British Columbia,}\\
\normalsize{Vancouver, BC V6T 1Z1, Canada}\\
\normalsize{$^{8}$LPCE / CNRS, F-45071 Orleans cedex 2, France}\\
\normalsize{$^{9}$Columbia Astrophysics Laboratory, Columbia University, New York, NY 10027, USA}\\
\normalsize{$^{10}$INAF - Osservatorio Astronomico di Cagliari, Poggio dei Pini, 09012 Capoterra, Italy}\\
\\
\normalsize{$^\ast$To whom correspondence should be addressed; E-mail: bretonr@physics.mcgill.ca.}
}

\date{}

\begin{document} 


\baselineskip24pt


\maketitle


\begin{sciabstract}
  The double pulsar PSR~J0737$-$3039A/B consists of two neutron stars in a
  highly relativistic orbit that displays a roughly 30-second eclipse when 
  pulsar A passes behind pulsar B. Describing this eclipse of pulsar A as due 
  to absorption occurring in the magnetosphere of pulsar B, we successfully 
  use a simple geometric model to characterize the observed changing eclipse 
  morphology and to measure the relativistic precession of pulsar B's spin 
  axis around the total orbital angular momentum. This provides a test of 
  general relativity and alternative theories of gravity in the strong-field 
  regime. Our measured relativistic spin precession rate of 
  $4.77^{+0.66}_{-0.65}\,^{\circ}\rm{yr}^{-1}$ (68\% confidence level) is 
  consistent with that predicted by general relativity within an uncertainty 
  of 13\%.
\end{sciabstract}


Spin is a fundamental property of most astrophysical bodies, making the study
of its gravitational interaction an important challenge \cite{wil01}. Spin
interaction manifests itself in different forms. For instance, we expect the
spin of a compact rotating body in a binary system with another compact
companion to couple gravitationally with the orbital angular momentum
(relativistic spin-orbit coupling) and also with the spin of this companion
(relativistic spin-spin coupling) \cite{foo1,oco74}. Observing such phenomena
provides important tests for theories of gravity, because every successful
theory must be able to describe the couplings and to predict their 
observational consequences. In a binary system consisting of compact 
objects such neutron stars, one can generally consider the spin-orbit 
contribution acting on each body to dominate greatly the 
spin-spin contribution. This interaction results in a precession of the 
bodies' spin axis around the orbital angular momentum of the system, 
behavior we refer to as {\em relativistic spin precession}.

While relativistic spin precession is well studied theoretically in general 
relativity (GR), the same is not true of alternative theories of gravity and
hence, quantitative predictions of deviations from GR spin precession do not 
yet exist \cite{dt92}. For instance, it is expected that in alternative theories 
relativistic spin precession may also depends on strong self-gravitational 
effects, i.e.~the actual precession may depend on the structure of a gravitating 
body \cite{dt92}. In the weak gravitational fields encountered in the solar 
system, these strong-field effects generally cannot be detected 
\cite{de92a,de92b,de96a}. Measurements in the strong-field regime near massive 
and compact bodies such as neutron stars and black holes are required. 
Relativistic spin precession has been observed in some binary pulsars 
[e.g. \cite{wrt89,kra98,hbo05}], but it has usually only provided a 
qualitative confirmation of the effect. Recently, the binary pulsar 
PSR~B1534+12 has allowed the first quantitative measurement of this effect 
in a strong field, and although the spin precession rate was measured to low
precision, it was consistent with the predictions of GR \cite{sta04}.

Here we report a precision measurement of relativistic spin precession using 
eclipses observed in the double pulsar \cite{bdp+03,lbk+04}. This measurement, 
combined with observational access to both pulsar orbits in this system, allows 
us to constrain quantitatively relativistic spin precession in the strong-field 
regime within a general class of gravitational theories that includes GR.

PSR~J0737$-$3039A/B consists of two neutron stars, both visible as radio pulsars,
in a relativistic 2.45-hour orbit \cite{bdp+03,lbk+04}. High-precision timing
of the pulsars, having spin periods of 23 ms and 2.8 s (hereafter called
pulsars A and B, respectively), has already proven to be the most stringent
test-bed for GR in the strong-field regime \cite{ksm+06}, and enables
four independent timing tests of gravity, more than any other binary system.

The orbital inclination of the double pulsar system is such that we observe the 
system almost perfectly edge-on. This coincidence causes pulsar A 
to be eclipsed by pulsar B at pulsar A's superior conjunction\cite{lbk+04}. 
The modestly frequency-dependent eclipse duration, about 30 s, corresponds to a
region extending $\sim$1.5$\times 10^7$\,m \cite{krb+04}. The light curve
of pulsar A during its eclipse shows flux modulations that are spaced by half
or integer numbers of pulsar B's rotational period \cite{mll+04}. This indicates 
that the material responsible for the eclipse corotates with pulsar B. The 
relative orbital motions of the two pulsars and the rotation of pulsar B thus 
allow a probe of different regions of pulsar B's magnetosphere in a plane 
containing the line of sight and the orbital motion.

Synchrotron resonance with relativistic electrons is the most likely mechanism
for efficient absorption of radio emission over a wide range of
frequencies. In the model proposed by Lyutikov and Thompson \cite{lt05},
this absorbing plasma corotates with pulsar B and is confined within the closed 
field lines of a magnetic dipole truncated by the relativistic wind of pulsar A. 
The dipole magnetic moment vector makes an angle $\alpha $ with respect to the 
spin axis of pulsar B, whose orientation in space can be described by two angles: 
the colatitude of the spin axis with respect to the total angular momentum of 
the system, $\theta $, and the longitude of the spin axis, $\phi $ (see Fig.~1 
for an illustration of the system geometry). Additional parameters 
characterizing the plasma opacity, $\mu$, the truncation radius of the 
magnetosphere, $R_{\rm mag}$, and the relative position of pulsar A with 
respect to the projected magnetosphere of pulsar B, $z_0$, are also included 
in the model \cite{lt05}.

We monitored the double pulsar from December 2003 to November 2007 using the
Green Bank Telescope in West Virginia; most of the data were acquired as part
of the timing observations reported in \cite{ksm+06}. The data
used for our analysis were taken at 820\,MHz with the SPIGOT instrument
\cite{kel+05}, which provides 1024 frequency channels across a 50\,MHz bandwidth. 
Data for a total of 63 eclipses of pulsar A were collected over the 4-year period, 
with many obtained during semi-annual concentrated observing campaigns. We 
dedispersed each eclipse data set by adding time shifts to frequency channels 
in order to compensate for the frequency-dependent travel time of radio waves 
in the ionized interstellar medium and we then folded them at the predicted 
spin period of pulsar A using the pulsar analysis packages \texttt{PRESTO} 
\cite{rem02} and \texttt{SIGPROC} \cite{sig} (see \cite{ksm+06} for details 
about the radio timing). Next, we extracted the relative pulsed flux density 
of pulsar A by fitting each folded interval for the amplitude of 
a high signal-to-noise ratio pulse profile template made from the integrated 
pulse observed during the several-hour observation that includes each 
eclipse. Finally, we normalized the flux densities so the average level outside 
the eclipse region corresponded to unity. We chose the time resolution of 
our eclipse light curves to equal, on average, four individual pulses of 
pulsar A ($\sim$91\,ms).

In addition to the flux density, we determined the orbital phase and the 
spin phase of pulsar B corresponding to each data point of our time 
series. Orbital phases were derived from the ephemeris published in
\cite{ksm+06}. Spin phases were empirically measured from data folded at 
the predicted period of pulsar B in a way similar to that described above 
for pulsar A. Over the four-year monitoring campaign, we found significant 
changes in pulsar B's pulse profile, likely due to the precession of its 
spin axis, which were also reported in \cite{bpm+05}. Around 2003, the 
average pulse profile was unimodal, resembling a Gaussian function. It 
evolved such that by 2007, it displayed two narrow peaks. Using the pulse 
peak maximum as a fiducial reference point is certainly not appropriate. 
We find, however, that the unimodal profile gradually became wider and 
then started to form a gap near the center of its peak. Since then, the 
outer edges of the pulse profile have not significantly changed but the 
gap evolved such that two peaks are now visible. This lets us presume that 
the underlying average profile is reminiscent of a Gaussian-like profile 
to which some ``absorption" feature has been superimposed near the 
center, leaving a narrow peak on each side. We therefore defined the 
fiducial reference point to lie at the center of the unimodal ``envelope" 
that we reconstructed from the first ten Fourier bins of the pulse 
profile, which contains 512 bins in total (see Fig.~S2 of the Supporting
Online Material for an illustration of the pulse profile evolution).

We implemented the eclipse modeling of our data in two steps: the fitting of
individual eclipse profiles and the search for evolution of the geometry of
pulsar B. We first searched the full phase space to identify best-fit values 
of six parameters (see Supporting Online Material for more details). Then, we 
reduced the number of free parameters to the subset ($\theta$, $\phi$, 
$\alpha$) describing the orientation of pulsar B's spin and magnetic axes 
by fixing the other parameters to their best-fit values: $\mu = 2$, 
$R_{\rm mag} = 1.29^\circ$ (projected value in terms of orbital phase) and 
$z_0/R_{\rm mag} = -0.543$ (see Fig.~1). Finally, we performed a 
high-resolution mapping of the likelihood of this subspace in order to 
investigate subtle changes in the geometry. Lyutikov and Thompson 
\cite{lt05} predicted that such changes, due to relativistic spin 
precession, could affect the eclipse light curve. In principle, 
relativistic spin precession of pulsar B's spin axis 
around the total angular momentum should induce a secular change of the 
longitude of the spin axis, $\phi $, while the magnetic inclination, 
$\alpha $, and the colatitude of the spin axis, $\theta $, are expected 
to remain fixed over time. Indeed, from model fitting, we find no 
significant time evolution of $\alpha $ and $\theta $, whereas $\phi $ does
change. Because of correlation between the parameters, we jointly
evaluated the best-fit geometry of pulsar B using a time-dependent model in
which $\alpha=\alpha_0 $ and $\theta=\theta_0 $ are constants, and $\phi $ varies
linearly with time, i.e. $\phi=\phi_0 - \Omega_{\rm B} t$, where $\Omega_{\rm B}$ is
the rate of change of pulsar B's spin axis longitude and the epoch of 
$\phi = \phi_0$ is May 2, 2006 (MJD 53857). Figure~2
shows the time evolution of the parameters and the fit derived from this joint
time-dependent model (Table~\ref{t:fit_result}). The precession rate 
$\Omega_{\rm B} $ of $4.77^{+0.66}_{-0.65}\,^{\circ}\rm{yr}^{-1}$\cite{foo8} 
agrees with the precession rate predicted by GR \cite{bo75}, 
$5.0734 \pm 0.0007\,^{\circ}\rm{yr}^{-1}$ \cite{foo16}, within an uncertainty 
of 13\% (68\% confidence level).

This relatively simple model \cite{lt05} is able to reproduce the complex 
phenomenology of the eclipses (see Fig.~3 and Movie~S1 in Supporting Online
Material) except at the eclipse boundaries where slight magnetospheric 
distortions or variations in plasma density are likely to occur. Fits 
including the egress generally are poor in the central region where we 
observe narrow modulation features, which are critical for determining 
pulsar B's geometry. For this reason, we excluded the egress from the fits, 
using orbital phases between $-1.0$ and $0.75^\circ$ (see Fig.~3). We 
accounted for systematics introduced by the choice of the region to fit 
in the priors of our Bayesian model (see Supporting Online Material). 
This improved the fit of the model throughout the center region of the 
eclipse while still producing qualitatively good predictions near the 
eclipse egress. The overall success of the model implies that the geometry 
of pulsar B's magnetosphere is accurately described as predominantly dipolar; 
a pure quadrupole, for instance, does not reproduce the observed light curves. 
Although the model does not exclude the possibility that higher-order 
multipole components may exist close to the surface of pulsar B, our 
modeling supports the conclusions \cite{lt05} that these eclipses yield
direct empirical evidence supporting the long-standing assumption 
that pulsars have mainly dipolar magnetic fields far from their surface.

The direct outcome from modeling the eclipse profile evolution is a measurement 
of the effect of relativistic spin precession (see Movie~S2 in Supporting Online
Material for an illustration of the time evolution of the eclipse). We can 
use the inferred precession rate to test GR (see Fig.~4) and to further
constrain alternative theories of gravity and the strong-field aspects of
relativistic spin precession. We use the generic class
of relativistic theories that are fully conservative (Lorentz-invariant) and
based on a Lagrangian, as introduced by Damour \& Taylor \cite{dt92}.
In this way we can study the constraints of our observations on theories of 
gravity by describing the spin-orbit interaction within a specific theory by 
coupling functions appearing in the corresponding part of the Lagrangian. 
In this framework, we can write the precession rate of pulsar B in a general 
form, $\Omega_{\rm B}= \sigma_B L/a^3_R (1-e^2)^{3/2}$ where $L$ is the 
orbital angular momentum of the system, $a_R$ is the semimajor axis of the 
relative orbit between the pulsars, $e$ the eccentricity of the orbit and 
$\sigma_B$ is a generic strong-field spin-orbit coupling constant. Since $L$ and 
$a_R$ are not directly measurable, it is more convenient to write the above
expression using observable Keplerian and post-Keplerian parameters. While 
alternative forms generally involve a mixture of gravitational theory-dependent 
terms, the particular choice $\Omega_{\rm B} = \frac{x_A x_B}{s^2} \times 
\frac{n^3}{1 - e^2} \times \frac{c^2 \sigma_B}{\cal G}$ is the only one that does 
not incorporate further theoretical terms other than the spin-orbit coupling 
constant, $\sigma_B$, the speed of light, $c$, and a generalized gravitational 
constant for the interaction between the two pulsars, ${\cal G}$. In this 
expression, the Keplerian parameters $e$ and $n=2\pi/P_b$, the angular orbital 
frequency, are easily measurable for any binary system. On the other hand, the 
post-Keplerian Shapiro delay shape parameter $s$, equivalent to the sine of
the orbital inclination angle \cite{dt92}, requires relatively edge-on orbits to
be observed. Measurement of the projected semi-major axes of the two orbits 
\cite{foo12}, $x_A$ and $x_B$, found in the above equation, necessitates that
each body must be timeable. Therefore, the 
double pulsar is the only relativistic binary system that allows a direct 
constraint on the spin-orbit coupling in general theories of gravity. Using the 
inferred precession rate of $\Omega_{\rm B}=4.77^{+0.66}_{-0.65}\,^{\circ}$ 
yr$^{-1}$, we derive $\left(\frac{c^2 \sigma_B}{\cal G}\right) = 3.38^{+0.49}_{-0.46}$. 
Every successful theory of gravity in the given generic framework must predict 
this value --- these observations provide a strong-field 
test of gravity that complements and goes beyond the weak-field tests of relativistic
spin precession \cite{oco08}. In GR, we expect to measure
$\left(\frac{c^2 \sigma_B}{\cal G}\right)_{\rm GR} = 2 + \frac{3}{2}\frac{m_A}{m_B} =
3.60677 \pm 0.00035$, where we have used the masses determined from the precisely 
observed orbital precession and the Shapiro delay shape parameter under the 
assumption that GR is correct \cite{ksm+06}. Comparing the observed value 
with GR's predictions, we find $\left( \frac{c^2 \sigma_B}{\cal G}
\right)_{\rm obs} / \left(\frac{c^2 \sigma_B}{\cal G}\right)_{\rm GR} =
0.94 \pm 0.13$. Hence, GR passes this test of relativistic 
spin precession in a strong-field regime, confirming, within uncertainties, 
GR's effacement property of gravity even for spinning bodies, i.e.~the notion 
that strong internal gravitational fields do not prevent a compact rotating 
body from behaving just like a spinning test particle in an external weak 
field \cite{dam87}.

The spin precession rate, as well as the timing parameters entering in the 
calculation of $\left( \frac{c^2 \sigma_B}{\cal G}\right)$, are all 
independent of the assumed theory of gravity.  If the main contribution 
limiting the precision of this new strong-field test comes from the inferred
spin precession rate, we expect that the statistical uncertainty should 
decrease significantly with time, roughly as the square of the monitoring baseline 
for similar quantity and quality of eclipse data. The contribution of systematics 
to the error budget should also decrease, but its functional time dependence is 
difficult to estimate. Although the orbital and spin phases of pulsar B 
are input variables to the eclipse model, our ability to determine the orientation 
of pulsar B in space does not require the degree of high-precision timing needed 
for measurement of post-Keplerian parameters; evaluating spin phases to the 
percent-level, for instance, is sufficient. Therefore, the intrinsic correctness 
of the model and its ability to reproduce future changes in the eclipse profile 
due to evolution of the geometry are the most likely limitations to improving the 
quality of this test of gravity, at least until the measured precession rate reaches 
a precision comparable with the timing parameters involved in the calculation of 
$\left(\frac{c^2 \sigma_B}{\cal G}\right)$. Better eclipse modeling could be achieved
from more sensitive observations and thus new generation radio telescopes such as the 
proposed Square Kilometer Array could help make important progress. Pulsar A does 
not show evidence of precession \cite{mkp+05,fsk+08} likely because its spin axis 
is aligned with the orbital angular momentum; it should therefore always remain 
visible, thus allowing long-term monitoring of its eclipses. Pulsar B, however, 
could disappear if spin precession causes its radio beam to miss our line of 
sight \cite{bpm+05}. In this event, we would need to find a way to circumvent the 
lack of observable spin phases for pulsar B, which are necessary to the eclipse 
fitting.

\nocite{foo9}





\clearpage

{\bf Fig.~1} Schematic view of the double pulsar system showing the important parameters for the modeling of pulsar A's eclipse (dimensions and angles are not to scale). Pulsar B is located at the origin of the cartesian coordinate system while the projected orbital motion of pulsar A during its eclipse is parallel to the $y$ axis at a constant $z_0$ as seen from Earth, which is located toward the positive $x$ axis. Note that since the orbital inclination is almost perfectly edge-on \cite{ksm+06}, we can approximate the $z$ axis to be coincident with the orbital angular momentum. The spin axis of pulsar B, whose spatial orientation is described by $\theta$ and $\phi$, is represented by the $\Omega$ vector. The magnetic axis of pulsar B corresponds to the $\mu$ vector and makes an angle $\alpha$ with respect to $\Omega$. Finally, the absorbing region of the dipolar magnetosphere of pulsar B, truncated at radius $R_{\rm mag}$, is shown as a shaded red region.

{\bf Fig.~2} Evolution of pulsar B's geometry as a function of time. The marginalized posterior probability distribution of the magnetic inclination ($\alpha$), the colatitude of the spin axis ($\theta$) and the longitude of the spin axis ($\phi$) of pulsar B are shown from top to bottom, respectively. For each data point, the circle represents the median value of the posterior probability density while the box and the bar indicate the 1$\sigma$ and 3$\sigma$ confidence intervals, respectively. The gray regions are the 3$\sigma$ confidence regions derived from the joint time-dependent model fitting. Note that for clarity, multiple eclipses are displayed as single data points when observed over an interval of about a week.

{\bf Fig.~3} Average eclipse profile of pulsar A consisting of eight eclipses observed at 820\,MHz over a five-day period around April 11, 2007 (black line) along with a model eclipse profile (red dashed line). The relative pulsed flux density of pulsar A is normalized so the average level outside the eclipse region is unity. The resolution of each data point is $\sim$91\,ms while 1$^\circ$ in orbital phase corresponds to 24.5\,s. Note that near orbital phase 0.0 the spikes are separated by the spin period of pulsar B.

{\bf Fig.~4} Mass-mass diagram illustrating the present tests constraining general relativity in the double pulsar system. The inset shows an expanded view of the region where the lines intersect. If general relativity is the correct theory of gravity, all lines should intersect at common values of masses. The mass ratio ($R = x_B/x_A$) and five post-Keplerian parameters ($s$ and $r$, Shapiro delay `shape' and `range'; $\dot \omega$, periastron advance; $\dot P_b$, orbital period decay due to the emission of gravitational waves; and $\gamma$, gravitational redshift and time dilation) were reported in \cite{ksm+06}. Shaded regions are unphysical solutions since $\sin i \le 1$, where $i$ is the orbital inclination. In addition to allowing a test of the strong-field parameter $\left(\frac{c^2 \sigma_B}{\cal G}\right)$, the spin precession rate of pulsar B, $\Omega_{\rm B}$, yields a new constraint on the mass-mass diagram.

\clearpage

\begin{table}
\begin{tabular}{lcccc}
\hline
Parameter & Mean & Median & $68.2\%$ Confidence & $99.7\%$ Confidence \\ 
\hline
$\alpha_0 $ & 70.92$^{\circ}$ & 70.94$^{\circ}$ & [70.49, 71.31]$^{\circ}$ & [69.68, 72.13]$^{\circ}$ \\
$\theta_0 $ & 130.02$^{\circ}$ & 130.02$^{\circ}$ & [129.58, 130.44]$^{\circ}$ & [128.79, 131.37]$^{\circ}$ \\
$\phi_0 $ & 51.21$^{\circ}$ & 51.20$^{\circ}$ & [50.39, 52.03]$^{\circ}$ & [48.80, 53.72]$^{\circ}$ \\
$\Omega_{\rm B} $ & 4.77$^{\circ}$yr$^{-1}$ & 4.76$^{\circ}$yr$^{-1}$ & [4.12, 5.43]$^{\circ}$yr$^{-1}$ & [2.89, 6.90]$^{\circ}$yr$^{-1}$ \\
\hline
\end{tabular}
\caption{Geometrical parameters of pulsar B derived from the eclipse model fitting. Note that the presented values include priors related to systematic uncertainties. The epoch of $\phi = \phi_0$ is May 2, 2006 (MJD 53857).}
\label{t:fit_result}
\end{table}

\clearpage
\centerline{\psfig{file=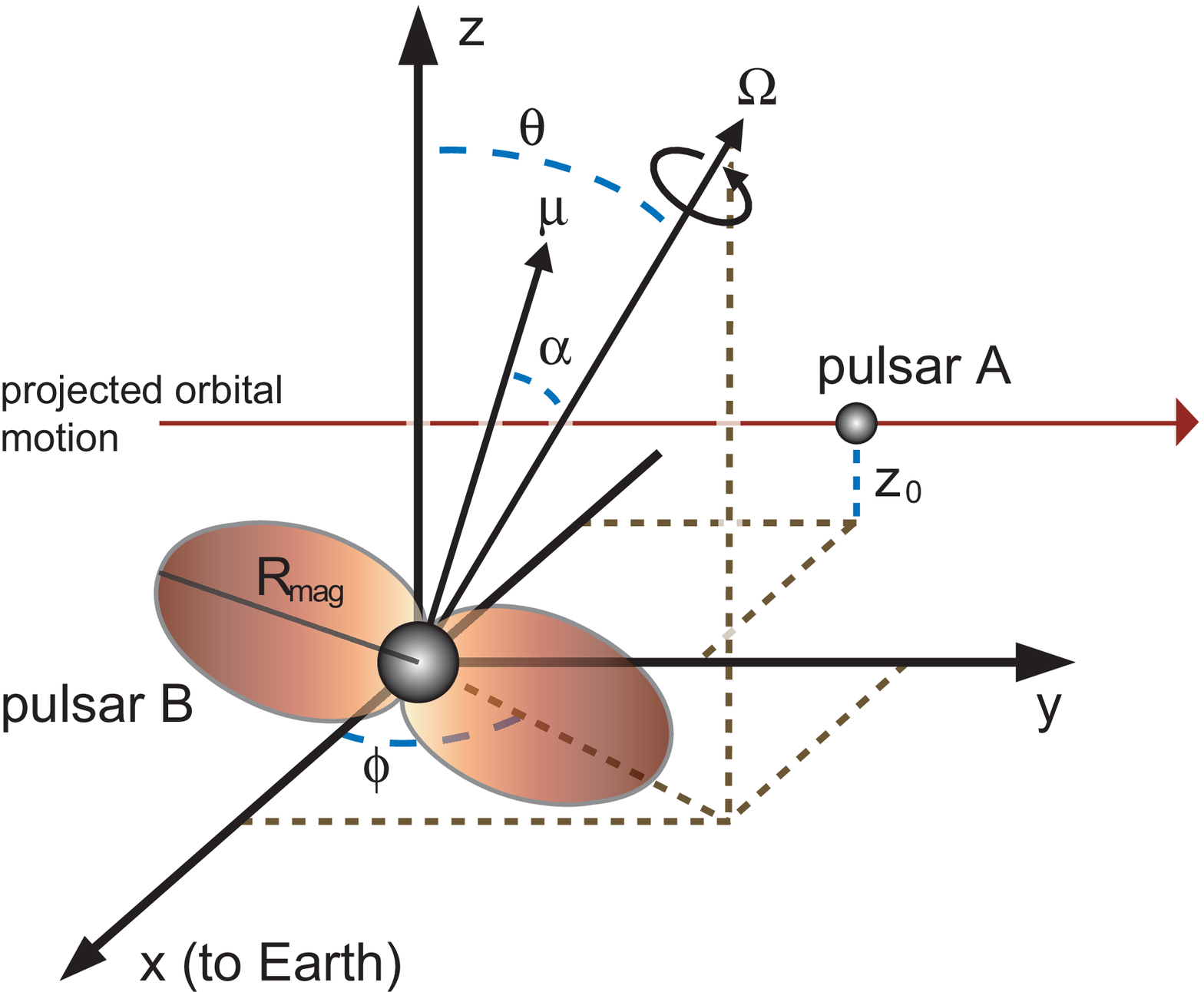,width=15cm}}
\centerline{\\Fig.~1}

\clearpage
\centerline{\psfig{file=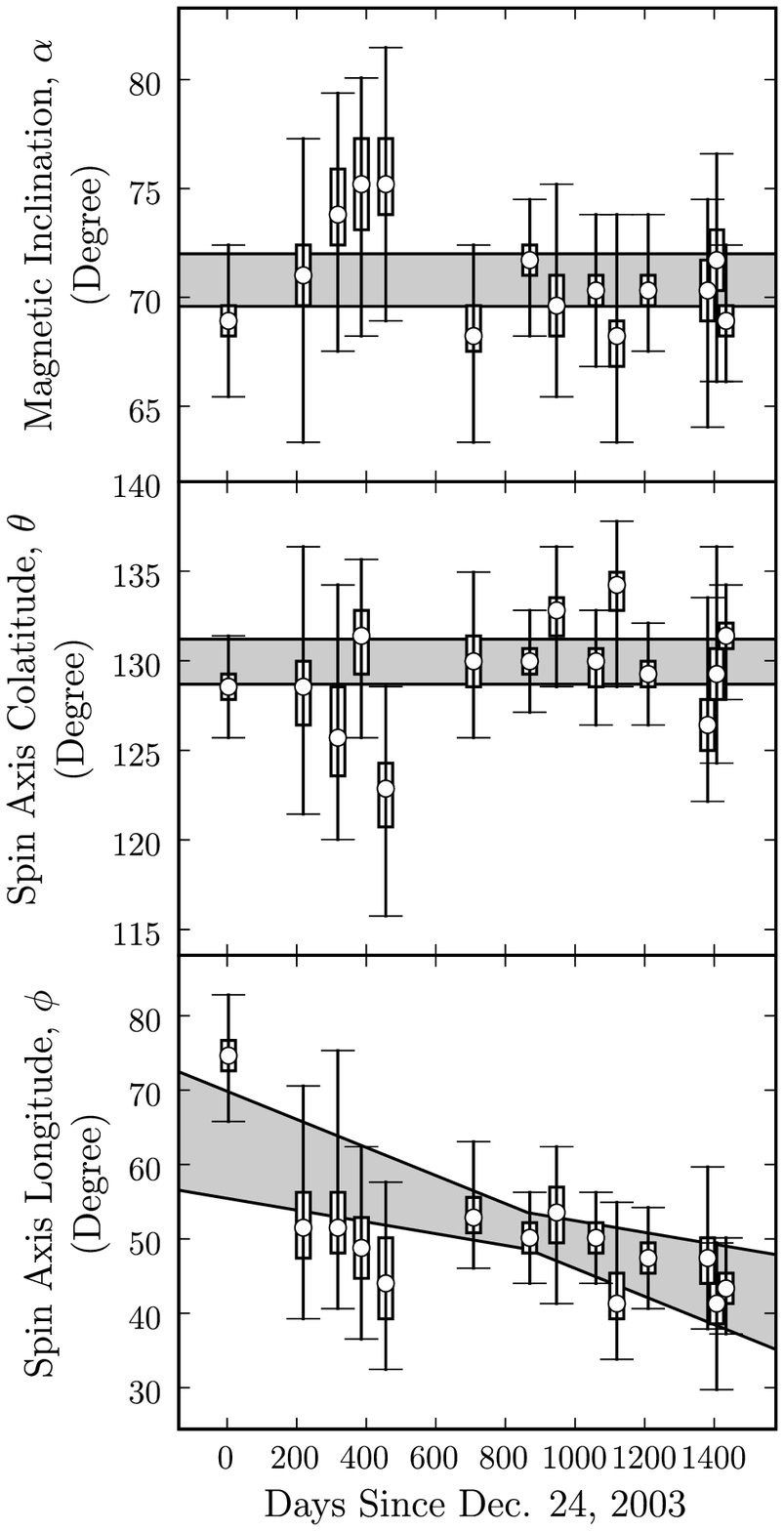,width=10cm}}
\centerline{\\Fig.~2}

\clearpage
\centerline{\psfig{file=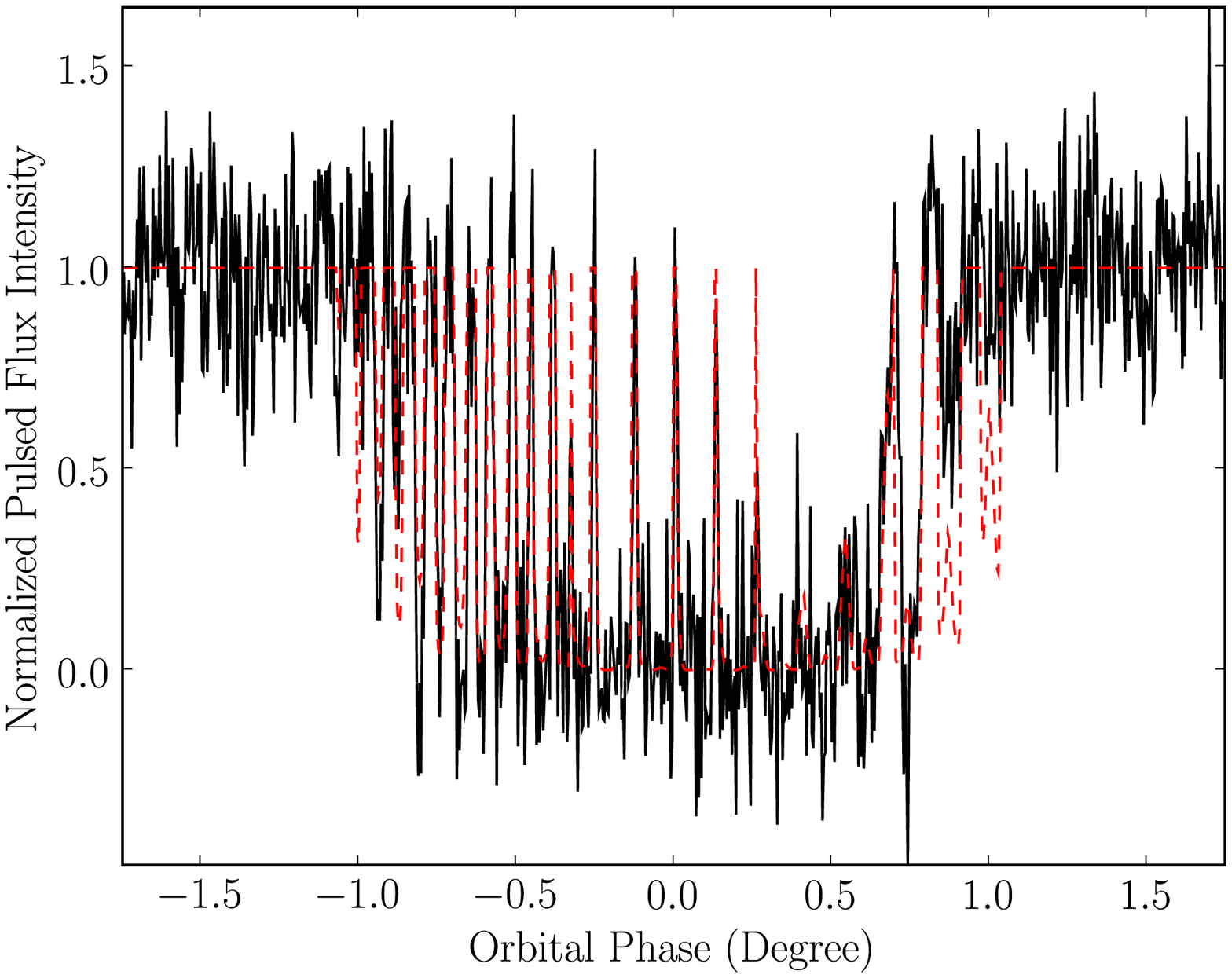,width=17cm}}
\centerline{\\Fig.~3}

\clearpage
\centerline{\psfig{file=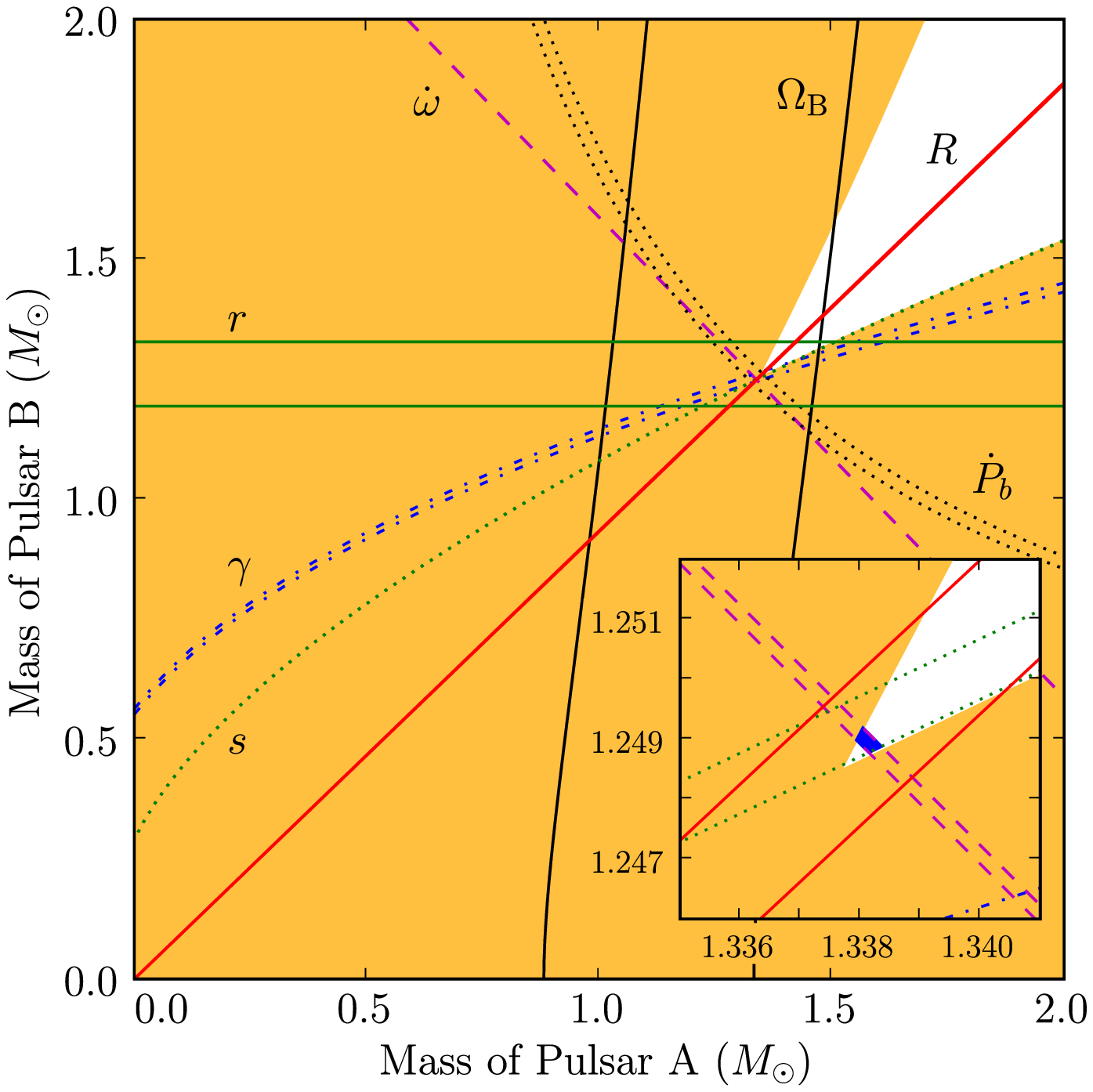,width=15cm}}
\centerline{\\Fig.~4}

\newpage

\centerline{\large \bf Supporting Online Material}

\section{Eclipse Model}

For the eclipse model we follow the prescription of \cite{lt05}, 
section 5.2, which proposes that the eclipses result from synchrotron
absorption of relativistic electrons trapped in the truncated dipolar magnetic
field of pulsar B. The intensity of the transmitted radio
emission from pulsar A is calculated from the optical depth of pulsar B's
magnetosphere, $\tau $. For a given data point in the light curve, the optical
depth is calculated as follows:
\begin{equation}
 \tau = \mu \int^{R_{\rm mag}}_{-R_{\rm mag}} \left( \frac{B \sin \kappa}{B_{\rm mag}} \right) d\left( \frac{x}{R_{\rm mag}} \right) \, ,
\end{equation}
where $\mu $ is a free parameter accounting for the characteristic optical
depth of the magnetosphere at 820\,MHz; $x$ is the radial position along our line 
of sight in units of $R_{\rm mag}$, the truncation radius of the dipole
magnetosphere; $B$ is the local magnetic dipolar field strength in units of 
$B_{\rm{mag}}$, the strength at the truncation radius; and $\kappa $ denotes the 
angle between our line of sight and the local direction of the magnetic field.
Beyond the truncation radius, the optical depth is assumed to be zero. Pulsar B
is at the origin of a coordinate system chosen so that $x$ is along our line of
sight, $y$ is in the plane of the sky along the projected orbital motion and $z$
is in the plane of the sky orthogonal to the $x-y$ plane (i.e. $z$ is parallel to 
the projected angular orbital momentum). In this coordinate system the apparent
motion of pulsar A is at a fixed value $z=z_0$. The orientation of the pulsar B 
spin axis can  be described by two angles, $\theta $ the colatitude 
of the spin axis with respect to the $z$ axis, and $\phi $ the longitude of the
spin axis with respect to the $x-z$ plane. Finally, the inclination of the
magnetic axis with respect to the spin axis of pulsar B is $\alpha $ (see Fig.~1
in the main article for a schematic view of the double pulsar showing the important 
model parameters).

Relativistic spin precession is expected to cause the spin
angular momentum vector of pulsar B to precess around the total angular
momentum of the system, which we take to be the same as the orbital angular
momentum since the spin contribution is negligible ($\sim$0.0001\%). The time 
evolution of the pulsar spin axis in a coordinate system aligned with the orbital 
angular momentum is given by:
\begin{eqnarray}
 \label{eqn:model1}
 \delta &=& \delta_0 \, , \\
 \label{eqn:model2}
 \phi_{\rm so} &=& \phi_{{\rm so}_0} - \Omega_{\rm B} t \, ,
\end{eqnarray}
where $\delta$ and $\phi_{\rm so}$ are the colatitude and the longitude of the
spin axis with respect to the total angular momentum, respectively
\cite{dt92}. They are related to our coordinate system in the following way:
\begin{eqnarray}
 \cos\theta &=& \cos(90^{\circ}-i) \cos\delta - \sin(90^{\circ}-i) \sin\delta \cos\phi_{\rm so} \, , \\
 \sin\phi &=& \frac{\sin\delta \sin\phi_{\rm so}}{\sin\theta} \,.
\end{eqnarray}

Because we observe the system almost perfectly edge-on ($i \approx
90^{\circ}$) the $z$ axis and the total angular momentum are almost perfectly
coincident. Hence, we can make the approximation $\theta \approx \delta$, and
$\phi \approx \phi_{\rm so}$. Therefore, the evolution of the spin geometry
given by equations \ref{eqn:model1} and \ref{eqn:model2} holds in our
coordinate system as well.

\section{Eclipse Model Fitting}

Our Bayesian fitting of the model to individual eclipse data involved the six
free parameters introduced above: $\mu $, $z_0$, $R_{\rm mag}$, $\alpha $,
$\theta $ and $\phi $ (see Fig.~1 in the main article for a schematic view of the
model geometry). We assumed flat priors for all parameters and evaluated
the joint posterior probability using a Markov Chain Monte Carlo \cite{gil95} for
the exploratory phase of the eclipse modelling. We coupled the Markov Chain Monte
Carlo to simulated annealing \cite{kgv83} in the early stage of the chain to boost 
up convergence and explore the parameter space more efficiently. We ran between
three and five independent Markov Chains for each fit in order to check for
consistency in the results. With this fitting optimization technique, we explored
the full parameter sub-space corresponding to the geometric orientation of pulsar 
B's magnetosphere (i.e. $\alpha $, $\theta $ and $\phi $) as well as enough of the
other parameters to identify the best-fit values.

A quantitative treatment of changes in the eclipses requires incorporation
of the eclipse model in a framework accounting for parameter
evolution. Except for $\phi $, we do not find any significant secular
evolution of the model parameters from their marginal posterior
probability \cite{foo7}. Indeed, theoretical considerations let us presume that only
$\phi $ should change over time due to relativistic spin precession. The Markov
Chain Monte Carlo technique poorly samples regions of low probability and so we
fixed $\mu = 2$, $R_{\rm mag} = 1.29^\circ$ (projected value in terms of orbital 
phase) and $z_0/R_{\rm mag} = -0.543$, their best-fit values, before making a
deeper investigation of pulsar B's spin axis evolution. Although $\theta $ and
$\alpha $ are not expected to vary, we did not fix them as they are highly covariant
with $\phi $, as opposed to the other parameters, but also because they
contain valuable information about pulsar B's geometry. Then, we completed our
analysis of the eclipse evolution using a high resolution grid to map the
now-reduced parameter space likelihood, containing three free
parameters. Following equations \ref{eqn:model1} and \ref{eqn:model2}, this
analysis implements a Bayesian joint fit that assumes $\alpha = \alpha_0 $ and
$\theta = \theta_0 $ are constant, and that $\phi $ linearly varies with time,
i.e. $\phi = \phi_0 - \Omega_{\rm B} t$, where $\Omega_{\rm B} $ is the precession
rate. The epoch of $\phi = \phi_0$ is May 2, 2006 (MJD 53857). Detailed results
from this fit showing different combinations of one-
and two-dimensional marginal posterior probabilities are presented in 
Fig.~S1. Values reported in Table 1 of the main article were derived from
these one-dimensional marginal posterior probability functions, assuming that
the integrated probabilities are unity.

\section{Analysis of Systematics}

We investigated the importance of systematics in the eclipse modeling and concluded that two main effects should contribute to increasing the total uncertainty in our best-fit geometric parameters, above the statistical value. First, we observe considerable changes in the pulse profile of pulsar B as shown in Fig.~S2. Since the spin phases of pulsar B are input data for the modeling, losing the fiducial reference to the neutron star surface will introduce additional error in the fitted eclipse parameters. While we do not require a measurement of the spin phases as accurate as for timing purposes, a few percent offset translates into slightly different geometrical parameters. The main effect of varying spin phases is to assign earlier or later rotational phases that mimic a slightly faster or slower precession rate. The pulse profile evolution of pulsar B is likely caused by the changing viewing geometry due to relavistic spin precession. Although it is not clear how pulsar B's pulse profile geometry is related to its surface, we are confident that the technique we used to determine the spin phases yields reliable results.

A second source of systematics arises from the choice of the eclipse region to include in the fit. Changes in the eclipse light curve due to relativistic spin precession are not uniform and the eclipse model tends to perform better toward the eclipse center than at the ingress or the egress. As opposed to the eclipse center, where our sight line to pulsar A goes deep inside and outside the magnetosphere of pulsar B as it rotates, our sight line only briefly intersects the edge of the magnetosphere at the beginning and the end of the eclipse. Therefore, local distortions of pulsar B's magnetic field or variations of the plasma density may give rise to a slight departure from our model. Indeed, we observe that fitting the whole eclipse does not generally provide qualitatively good fits. The narrow and periodic modulations in the eclipse center are very important markers for the geometric orientation of pulsar B but they tend to be misfitted because broader features in the egress region lead to larger variations of the goodness-of-fit. We find that excluding the egress more accurately fits the overall light curves, without sacrificing critical information derived from narrow modulations, while still qualitatively reproducing the egress. Therefore, we chose to fit the eclipse in the range $[-1.0^{\circ},0.75^{\circ}]$ centered around conjunction (see Fig.~3 of the main article for an example of a fitted eclipse light curve).

Determining the boundaries of the region to fit is arbitrary and hence we estimated how much dispersion in the best-fit values is induced by other choices of limits. We compared our actual choice, $[-1.0^{\circ},0.75^{\circ}]$, with the full eclipse, $[-1.0^{\circ},1.0^{\circ}]$, the eclipse center, $[-0.6^{\circ},0.6^{\circ}]$, and the extended center, $[-0.7^{\circ},0.7^{\circ}]$, fits. In a Bayesian framework, we can easily incorporate the effect of systematics as priors on the model parameters. For simplicity and because the functional form of the systematics is poorly defined we assume Gaussian priors. Therefore, we can recast the posterior probability distribution of our pre-systematics analysis work, for which we were assuming constant priors:
\begin{equation}
 p\left(\alpha,\theta,\phi | D\right) \propto {\cal L} \left(D | \alpha,\theta,\phi \right) \, ,
\end{equation}
as:
\begin{equation}
 p\left(\alpha,\theta,\phi | D\right) \propto
 	\int_{\alpha^{\prime}} {\cal N}(\alpha - \alpha^{\prime}, \sigma_{\alpha})
 	\int_{\theta^{\prime}} {\cal N}(\theta - \theta^{\prime}, \sigma_{\theta})
 	\int_{\phi^{\prime}} {\cal N}(\phi - \phi^{\prime}, \sigma_{\phi}) \,
 	{\cal L} \left( D | \alpha^{\prime},\theta^{\prime},\phi^{\prime} \right) \, d\alpha^{\prime} \, d\theta^{\prime} \, d\phi^{\prime}
 \, ,
\end{equation}
where ${\cal N}(\nu, \sigma_{\nu})$ is a Gaussian distribution of mean $\nu$ and standard deviation $\sigma_{\nu}$. The likelihood, ${\cal L} \left(D | \alpha,\theta,\phi \right)$, is defined as $\exp(-\chi^2_\nu / 2)$, with $\chi^2_\nu$ being the standard reduced chi-square. From the analysis of systematics due to the choice of the region to fit, along with the additional uncertainty in the spin phase of pulsar B due to the long-term pulse profile variations, we estimate that systematics contribute $\sigma_{\alpha} = 1^{\circ}$, $\sigma_{\theta} = 1^{\circ}$ and $\sigma_{\phi} = 2.0^{\circ}$. Note that incorporating Gaussian priors due to systematics has the effect of convolving the three-dimensional likelihood obtained from the eclipse fitting, ${\cal L} \left(D | \alpha,\theta,\phi \right)$, with a three-dimensional Gaussian. The analysis reported in this article includes these priors (this is particularly relevant for numbers quoted in Table 1 of the main article and for the marginalized posterior probability distributions presented in Fig.~S1 of this Supporting Online Material).


\begin{thebibliography}{10}
\bibitem{foo1}
The contribution from a classical quadrupolar moment is negligible for compact
  bodies.
\bibitem{foo8}
Unless otherwise stated, uncertainties are quoted at the 68\% confidence level.
\bibitem{foo16}
The uncertainty on the predicted GR spin precession rate arises because the
  value depends on the masses of the system, which are determined from two
  measured post-Keplerian parameters: the Shapiro delay s parameter and the
  advance of periastron $\dot\omega$.
\bibitem{foo12}
The projected semi-major axes are expressed in terms of light-travel time
  across the orbit.
\bibitem{foo7}
Marginalization is a common technique in Bayesian analysis that consists of
  integrating the joint posterior probability over the parameters that are not
  of interest so that it is reduced, in this case, to a one-dimensional
  probability for the parameter that we wish to consider \cite{gre05}.
\end{thebibliography}




\clearpage

{\bf Fig.~S1} Pulsar B's pulse profile consisting of the integrated flux at all orbital phases for each observation in our data set. Pulse profiles are normalized so the peak is unity and they are displayed with an incremental 1-unit vertical shift for clarity (the vertical axis does not show a linear time sequence). The vertical red dashed line marks the fiducial spin phase, which was determined by aligning the profiles using the first ten Fourier bins of the original 512-bin profile assuming that the two narrow peaks visible in the more recent data are the ``edges" of an underlying unimodal envelope reminiscent of the profile in the earlier observations. Note that the observation length and radio interference contamination slightly varies from epoch to epoch but the overall signal-to-noise ratio decreases in the latest observations due to pulsar B becoming weaker. MJD 52996 corresponds to Dec. 23, 2003, and MJD 54430 to Nov. 26, 2007.

{\bf Fig.~S2} One- and two-dimensional projections of the marginalized posterior probability distributions for the joint fit of the parameters' evolution. Black contours in two-dimensional maps are joint $1,2,3,4$ and $5\sigma $ confidence regions, with the red color being associated with a higher likelihood value. The epoch of $\phi = \phi_0$ is May 2, 2006 (MJD 53857). Note that these probability distributions include priors related to systematic uncertainties (see Section 3 of this Supporting Online Material).

\clearpage

{\bf Movie~S1}\footnote{Movies, as well as the published version of this paper, can be found at \url{http://www.sciencemag.org/cgi/content/full/sci;321/5885/104/DC1}.} The eclipses in the double pulsar PSR~J0737$-$3039A/B occur when pulsar A's projected orbital motion, represented by a gray circle moving on a black line, passes behind its companion, pulsar B. Radio emission from pulsar A is absorbed via synchrotron resonance with the plasma trapped in the closed field lines of the truncated dipolar magnetosphere of pulsar B, shown as a colored dipolar structure. Since pulsar B's magnetic dipole axis is misaligned with respect to its spin axis (represented by a diagonal rod), the opacity along our sight line to pulsar A varies as a function of pulsar B's spin phase. The theoretical light curve resulting from the eclipse animated in the upper panel is drawn as a black curve in the bottom panel and real eclipse data, observed with the Green Bank Telescope in April 2007, are overlaid in red.

{\bf Movie~S2}\footnotemark[\value{footnote}] Time-lapse animation displaying the evolution of pulsar B's geometry in the double pulsar PSR~J0737$-$3039A/B due to relativistic spin precession between January 2004 and January 2029. The truncated dipolar magnetosphere of pulsar B, shown as a colored dipolar structure, rotates about its spin axis, pictured as a diagonal rod. The apparent orbital motion of pulsar A during the eclipse corresponds to the horizontal black line intersecting pulsar B's magnetosphere. Relativistic spin precession is similar to the wobbling of a spinning top and induces a motion of the spin-axis orientation around the orbital angular momentum, which is vertical in this movie. The theoretical light curve corresponding to the eclipse animated in the upper panel is drawn in the lower panel. The angle $\phi$ corresponds to the longitude of the spin axis, with $0^\circ$ being the direction coincident with the line of sight.

\clearpage
\centerline{\psfig{file=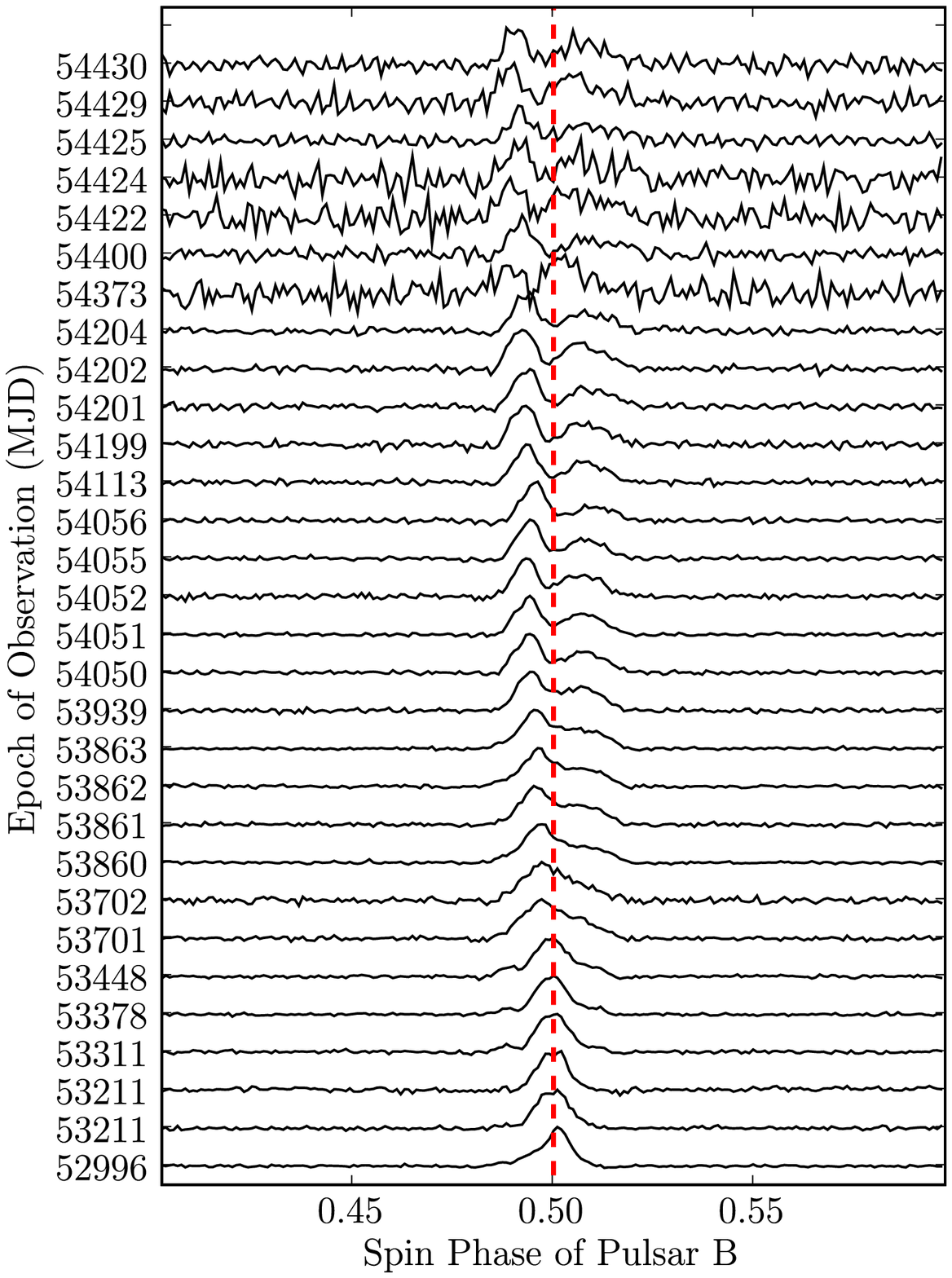,width=15cm}}
\centerline{\\Fig.~S1}

\clearpage
\centerline{\psfig{file=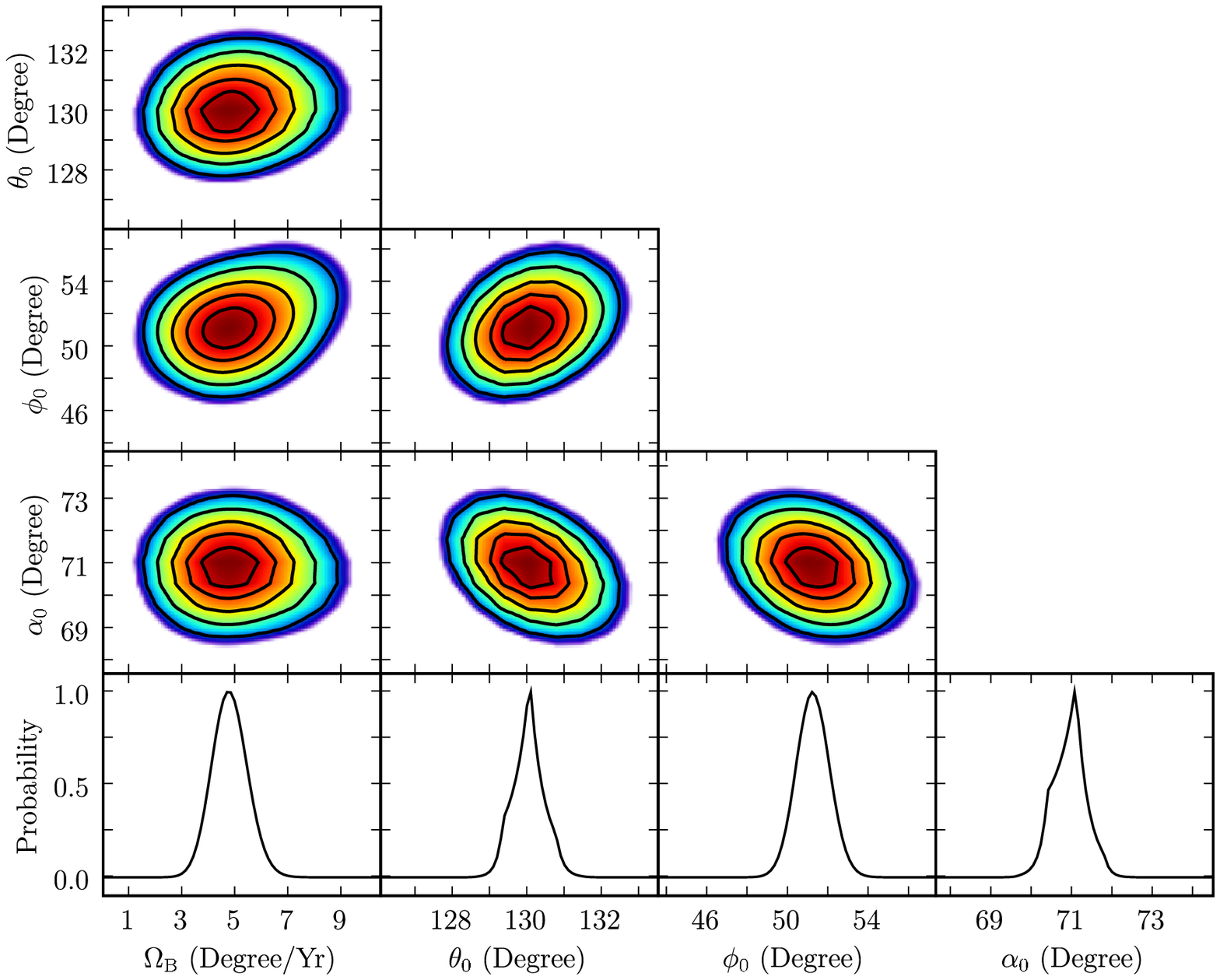,width=16cm}}
\centerline{\\Fig.~S2}

\end{document}